\RequirePackage{fix-cm}
\documentclass[a4paper,11pt]{article}

\usepackage{authblk}
\usepackage{amsfonts,amsmath,amssymb,amsthm}
\usepackage{algorithm}
\usepackage{paralist}
\usepackage{booktabs}
\usepackage{url}
\usepackage{hyperref}
\usepackage{graphicx}
\usepackage{microtype}
\usepackage{url}
\usepackage{hyperref}
\usepackage{subcaption}
\usepackage{tikz}
\usepackage{multirow}
\usepackage{wasysym}
\usetikzlibrary{positioning}
\usetikzlibrary{decorations.pathreplacing}
\newcommand{\N}{\mathbb{N}}
\newcommand{\ZZ}{\mathbb{Z}}
\newcommand{\F}{\mathbb{F}}

\theoremstyle{thmstyleone}%
\newtheorem{theorem}{Theorem}% 
\newtheorem{lemma}[theorem]{Lemma}% 
\theoremstyle{thmstyletwo}%
\newtheorem{resq}{Research Question}%
\theoremstyle{thmstylethree}%
\newtheorem{definition}{Definition}%

\tikzset{
    mybrace/.style={decorate,decoration={brace,aspect=#1}}
}

\providecommand{\keywords}[1]{\textbf{\textit{Keywords }} #1}

\begin{document}

\title{A classification of S-boxes generated by Orthogonal Cellular Automata}

\author[1]{Luca Mariot}
\author[2]{Luca Manzoni}
	
\affil[1]{{\small Semantics, Cybersecurity and Services Group, University of Twente, Drienerlolaan 5, 7511GG Enschede, The Netherlands} 
	
	{\small \texttt{l.mariot@utwente.nl}}}

\affil[2]{{\small Dipartimento di Matematica e Geoscienze, Università degli Studi di Trieste, Via Valerio 12/1, Trieste, 34127, Italy}

    {\small \texttt{lmanzoni@units.it}}}

\maketitle

\begin{abstract}
Most of the approaches published in the literature to construct S-boxes via Cellular Automata (CA) work by either iterating a finite CA for several time steps, or by a one-shot application of the global rule. The main characteristic that brings together these works is that they employ a single CA rule to define the vectorial Boolean function of the S-box. In this work, we explore a different direction for the design of S-boxes that leverages on Orthogonal CA (OCA), i.e. pairs of CA rules giving rise to orthogonal Latin squares. The motivation stands on the facts that an OCA pair already defines a bijective transformation, and moreover the orthogonality property of the resulting Latin squares ensures a minimum amount of diffusion. We exhaustively enumerate all S-boxes generated by OCA pairs of diameter $4 \le d \le 6$, and measure their nonlinearity. Interestingly, we observe that for $d=4$ and $d=5$ all S-boxes are linear, despite the underlying CA local rules being nonlinear. The smallest nonlinear S-boxes emerges for $d=6$, but their nonlinearity is still too low to be used in practice. Nonetheless, we unearth an interesting structure of linear OCA S-boxes, proving that their Linear Components Space (LCS) is itself the image of a linear CA, or equivalently a polynomial code. We finally classify all linear OCA S-boxes in terms of their generator polynomials.
\end{abstract}


\section{Introduction}
\label{sec:intro}
In cryptography, the design of symmetric ciphers is usually based on the key principles of \emph{confusion} and \emph{diffusion} set forth by Shannon in 1949~\cite{shannon49}. The confusion principle states that the relationship between the ciphertext and the secret key should be as complicated as possible. The diffusion principle, on the other hand, prescribes that the statistical structure of the plaintext should be spread as much as possible over the ciphertext.

These two principles are usually implemented through different designs approaches. In the context of block ciphers, one of the most widespread approaches is the \emph{Substitution-Permutation Network} (SPN)~\cite{stinson18}. For example, the {\sc Rijndael} cipher, which has been selected by the NIST for the the Advanced Encryption Standard (AES), is designed following the SPN paradigm~\cite{daemen20}. The idea behind SPN ciphers is that a fixed-length block of plaintext is first passed through a \emph{substitution layer}, which realizes the confusion principle, and then fed into a \emph{permutation layer}, that embodies the diffusion principle. In particular, the substitution layer is usually implemented with a series of small \emph{Substitution Boxes} (S-boxes), that are basically vectorial Boolean functions mapping $n$-bit input vectors to $n$-bit output vectors. The plaintext block is thus processed by chopping it into $n$-bit sub-blocks, each of which is transformed by the corresponding S-box. Usually, to save on the implementation cost the same S-box is used for all sub-blocks. 

The S-boxes used in the substitution layer of a SPN cipher need to satisfy a number of properties, both for security and implementation reasons. For example, to ensure that the ciphertext can be decrypted back to the plaintext, the S-box need to be \emph{bijective}. Further, the S-box must also have a high \emph{nonlinearity}, to withstand linear cryptanalysis attacks. This property is directly related to the confusion principle.

There are several ways to construct S-boxes that are both bijective and highly nonlinear, and most of the methods proposed in the literature employ algebraic constructions. For instance, {\sc Rijndael} uses an $8\times 8$ S-box that computes the inversion over the finite field $\F_{2^8}$, and which is applied in parallel over a 128-bit plaintext block. Other than the one used in {\sc Rijndael}, many other S-boxes of different sizes and defined by different operations have been considered in this research thread. The choice of a specific S-box mainly depends on the security and efficiency requirements for a particular cipher. For example, \emph{lightweight} ciphers such as {\sc Present}~\cite{bogdanov07} and {\sc Rectangle}~\cite{zhang15} employ small $4 \times 4$ S-boxes, since they are designed for very efficient hardware implementations.

An interesting approach for the design of strong S-boxes is based on \emph{Cellular Automata} (CA), which provide a good trade-off between security and efficiency. The efficiency advantage stems from the fact that CA are \emph{shift-invariant} functions, or equivalently the same local rule is applied in parallel at all output coordinates of the S-box. Thus, the design problem reduces to the choice of a good local rule that induces a globally invertible S-box with high nonlinearity. The {\sc Keccak} sponge construction~\cite{bertoni13}, which has been selected by the NIST in 2012 as the new SHA-3 standard for cryptographic hash functions~\cite{dworkin15}, is perhaps the best known example of a symmetric primitive that uses a cellular automaton to achieve confusion. In particular, the permutation in {\sc Keccak} employs a $5 \times 5$ S-box defined by the elementary CA~$\chi$, which corresponds to rule $210$ in Wolfram's numbering convention.

In the context of the literature related to CA-based S-boxes, a one-dimensional CA with periodic boundary conditions is usually seen as a particular kind of vectorial Boolean function, whose cryptographic properties need to be optimized. Most of the works in this research line consider the CA as a discrete dynamical system, which is iterated for multiple time steps to compute the output of the S-box~\cite{seredynski04,seredynski04a,marconi06,szaban08,oliveira10}. Other works study instead the S-box arising from a one-shot evaluation of the CA global rule, as in the case of {\sc Keccak}~\cite{bertoni06,picek17,mariot19}.

In this paper, we take on a different perspective to design S-boxes with CA, namely by employing results from the research line of \emph{orthogonal CA} (OCA). Two CA defined by bipermutive local rules are called orthogonal if they can be used to generate a pair of \emph{orthogonal Latin squares}~\cite{mariot18}. The motivation to employ OCA to design S-boxes is twofold: first, the superposition of two orthogonal Latin squares defines a permutation on the Cartesian product of the set of entries. Hence, the superposition of two OCA yields a bijective S-box. Second, orthogonal Latin squares are equivalent to $(2,2)-$\emph{multipermutations}, and thus they ensure a minimum amount of diffusion~\cite{vaudenay94}. Therefore, the research question addressed in this paper is the following: \emph{is it possible to use OCA to define S-boxes with high nonlinearity?} While the theory of linear OCA is already well-developed~\cite{mariot20}, we observe that much less is known about nonlinear OCA~\cite{mariot17}, which seem to be a necessary condition to construct highly nonlinear S-boxes from OCA.

This manuscript is an extended version of the conference paper ``\emph{On the Linear Components Space of S-boxes Generated by Orthogonal Cellular Automata.}'' presented at ACRI 2022~\cite{mariot22c}. There, the main results were as follows:
\begin{enumerate}
\item We exhaustively searched the set of OCA pairs of diameter $d=4$ and $d=5$, remarking that \emph{all of them generate linear S-boxes}. This ruled out the possibility of using the corresponding S-boxes in practice.
\item We remarked an interesting coding-theoretic structure in the Linear Component Space (LCS) of the S-boxes generated by OCA. Indeed, we empirically observed all these vector spaces are actually \emph{polynomial codes}. The interesting aspect is that the generator matrix of a polynomial code is itself the transition matrix of a linear CA.
\end{enumerate}

In this paper, we introduce the following new contributions with respect to the conference version:
\begin{itemize}
\item We extend our exhaustive search experiments up to OCA of diameter $d=6$, exploiting the combinatorial enumeration algorithm of~\cite{mariot17}. \emph{The results show that nonlinear OCA S-boxes indeed exist}. Unfortunately, the observed values of nonlinearity are still too low to grant the use of these S-boxes in the substitution layer of SPN ciphers.
\item We formally prove the conjecture formulated in our conference paper, i.e. that if the S-box generated by a pair of OCA is linear, then its LCS is always a polynomial code.
\item Finally, we provide a classification of all linear OCA S-boxes in terms of the generator polynomials of their LCS.
\end{itemize}

In perspective, the findings above are negative concerning the use of OCA for the design of S-boxes to be used in the confusion layers of block ciphers. However, the characterization of the LCS of linear S-boxes might give some insights for future research on the characterization of nonlinear OCA.

The rest of this paper is organized as follows. Section~\ref{sec:bg} covers all background definitions and concepts related to Boolean functions, S-boxes and cellular automata used throughout the paper. Section~\ref{sec:rel-work} gives a general overview of the literature concerning the design of S-boxes with CA. Section~\ref{sec:sboxes} describes our method to construct a S-box from a pair of OCA. Section~\ref{sec:search} reports the results of the exhaustive search experiments conducted for OCA pairs of diameter $4 \le d \le 6$. Section~\ref{sec:lcs} analyzes the LCS of linear OCA S-boxes, and provide a characterization in terms of their generator polynomials. Finally, Section~\ref{sec:outro} recaps the key contributions of the paper, and points out a few directions for future research on the subject.

\section{Background}
\label{sec:bg}
In this section, we introduce all basic definitions and results related to Boolean functions and cellular automata that we use in the remainder of the paper. The discussion here is far from complete: for a thorough treatment on Boolean functions and S-boxes, we refer the reader to Carlet's recent book~\cite{carlet21}. For orthogonal CA we follow the notation and terminology of~\cite{mariot18,mariot20}.

\subsection{Boolean Functions and S-boxes}
\label{subsec:boolfun}
In what follows, we denote by $\F_2 = \{0,1\}$ the finite field with two elements. The field operations correspond respectively to the XOR (denoted by $\oplus$) and logical AND (denoted by concatenation) of two elements. Given $n \in \N$, the $n$-dimensional vector space of all $n$-bit strings is denoted by $\F_2^n$. The sum between two vectors $x, y \in \F_2^n$ is defined as their bitwise XOR (and, slightly abusing notation, still denoted as $x \oplus y$). The multiplication of a vector $x \in \F_2^n$ by a scalar $a \in \F_2$ is the field multiplication of each coordinate of $x$ by $a$. In particular, this implies that two vectors $x,y \in \F_2^n$ are linearly independent if and only if $x \neq y$, and that a set of $k$ vectors in $\F_2^n$ is linearly independent if and only if each vector cannot be written as a bitwise XOR of a subset of the others. Further, the \emph{dot product} of two vectors $x, y \in \F_2^n$ is defined as $x \cdot y = \bigoplus_{i=1}^n x_iy_i$, while their \emph{Hamming distance} $d_H(x,y) = \#\{i: x_i \neq y_i\}$ is the number of coordinates where $x$ and $y$ disagree. The support of $x \in \F_2^n$ is the set of nonzero coordinates of $x$, that is, $supp(x) = \{i : x_i \neq 0\}$. The \emph{Hamming weight} of $x \in \F_2^n$ is the size of its support, i.e. $w_H(x) = |supp(x)|$. Equivalently, the Hamming weight of $x$ is the Hamming distance between $x$ and the null vector $\underbar{0}$, or the number of ones in $x$.

A Boolean function of $n$ variables is a mapping $f: \F_2^n \to \F_2$, and there are several ways to uniquely represent it. Here, we focus on the three representations that are most commonly used in cryptography, namely the truth table, the algebraic normal form and the Walsh transform. In what follows, we assume that the vectors of $\F_2^n$ are lexicographically ordered.

The \emph{truth table} of $f: \F_2^n \to \F_2$ is the $2^n$-bit vector $\Omega_f \in \F_2^{2^n}$ which specifies for each input vector $x \in \F_2^n$ the corresponding output value $f(x)$. The weight of a Boolean function is the Hamming weight of its truth table. In particular, a function $f: \F_2^n \to \F_2$ is called \emph{balanced} if $w_H(\Omega_f) = 2^{n-1}$, i.e. if its truth table has an equal number of zeros and ones. Balanced functions are usually sought in the design of combiner or filter stream ciphers, to avoid introducing any statistical bias in the ciphertext that might be exploited by an attacker.

Remarking that $x^2 = x$ for all elements $x \in \F_2$, the \emph{Algebraic Normal Form} (ANF) of $f$ is the multivariate polynomial in the quotient ring $\F_2[x_1,\cdots,x_n]/[x_1^2\oplus x_1, \cdots , x_n^2 \oplus x_n]$ defined as:
\begin{equation}
\label{eq:anf}
P_f(x) = \bigoplus_{u \in \F_2^n} a_u x^u = \bigoplus_{u \in F_2^n} a_u x_1^{u_1}x_2^{u_2}\ldots x_n^{u_n} \enspace ,
\end{equation}
where $a_u \in \F_2$ for all $u \in \F_2^n$. The coefficients $a_u \in \F_2^n$ in the 
ANF of $f$ can be obtained through the \emph{binary M\"{o}bius transform}:
\begin{equation}
\label{eq:anf}
a_u = \bigoplus_{x \in \F_2^n : supp(x) \subseteq supp(u)} f(x) \enspace .
\end{equation}
Notice that the binary M\"{o}bius transform is an \emph{involution}: therefore, one may retrieve the truth table of a function from its ANF coefficients by using the above formula, swapping $a$ with $f$ and $u$ with $x$. The \emph{algebraic degree} of $f$ corresponds to the size of the largest monomial occurring in its ANF; formally, this is defined as $deg(f) = \max_{u \in \F_2^n} \{w_H(u): u \neq 0\}$. Functions of degree $1$ are also called \emph{affine}, and an affine function is called \emph{linear} if $a_{\underline{0}} = 0$ (i.e., the ANF of $f$ does not have any constant term). The ANF of a linear function can be defined as a dot product $a\cdot x$, with $a \in \F_2^n$.

The \emph{Walsh transform} of $f: \F_2^n \to \F_2$ is the map $W_f: \F_2^n \to \ZZ$ defined for all $a \in \F_2^n$ as:
\begin{equation}
\label{eq:wt}
W_f(a) = \sum_{x \in \F_2^n} (-1)^{f(x) \oplus a\cdot x}, \enspace ,
\end{equation}
Intuitively, the coefficient $W_f(a)$ measures the \emph{correlation} between $f(x)$ and the linear function defined by $a\cdot x$. The \emph{nonlinearity} of a Boolean function $f: \F_2^n \to \F_2$ is the minimum Hamming distance of $f$ from the set of all $n$-variable affine functions. Using the Walsh transform, the nonlinearity can be computed as follows:
\begin{equation}
\label{eq:nl}
nl(f) = 2^{n-1} - \frac{1}{2} \max_{a \in \F_2^n}\left\{ |W_f(a)| \right\} \enspace .
\end{equation}
As a cryptographic criterion, the nonlinearity of Boolean functions used in stream and block ciphers should be as high as possible to withstand fast-correlation attacks and linear cryptanalysis, respectively.

\emph{Substitution Boxes} (S-boxes) are the vectorial generalization of Boolean functions. Given $n,m \in \N$, a $(n,m)$-function or S-box is a vectorial mapping of the form $F: \F_2^n \to \F_2^m$, which is defined by its \emph{coordinate functions}. For all $i \in \{1,\cdots, m\}$, the coordinate function $f_i: \F_2^n \to \F_2$ is a $n$-variable Boolean function that specifies the $i$-th output bit of the S-box $F$. The \emph{component functions} of $F: \F_2^{n} \to \F_2^{m}$ are defined as the nontrivial linear combinations of the coordinate functions of $F$. More precisely, given a vector $v \in \F_2^m \setminus \{\underbar{0}\}$, the corresponding component function $v\cdot F: \F_2^n \to \F_2$ of $F$ is defined as the dot product $v\cdot F(x)$, for all $x \in \F_2^n$.

In the rest of this work, we will be mostly interested in S-boxes where $n=m$, since these are the most commonly used in SPN ciphers. The cryptographic properties described above for Boolean functions are generalized to the vectorial case by using their coordinate and component functions. In particular, an S-box $F:\F_2^n \to \F_2$ is balanced if and only if all its component functions are balanced. Remark that balancedness in $(n,n)$-functions corresponds to bijectivity, and this explains why S-boxes in the SPN paradigm are sought balanced: if they are not, then decryption is not possible. The algebraic degree of an S-box, on the other hand, is defined as the \emph{maximum} algebraic degree among all its \emph{coordinate} functions. Finally, the nonlinearity of an S-box $F: \F_2^n \to \F_2$ is defined as the \emph{minimum} nonlinearity among all its \emph{component} functions. This means that a single linear component functions suffices to make the whole S-box linear.

Finally, remark that the set $\mathcal{L}_F = \{v \in \F_2^{m} \setminus\{\underline{0}\} : nl(v\cdot F) = 0\}$ of all linear component functions of an S-box $F$ is a subspace of $\F_2^m$. As a matter of fact, if two functions are affine, their sum must be affine too. We call $\mathcal{L}_F$ the \emph{linear components space} (LCS) of $F$, and we will use it in later sections to classify the S-boxes generated by orthogonal cellular automata.

\subsection{Orthogonal Cellular Automata}
\label{subsec:ca}
A \emph{Cellular Automaton} (CA) is a discrete computational model made of a regular lattice of \emph{cells}, also called a \emph{cellular array}. In what follows, we will focus only on one-dimensional CA, meaning that the cellular array is basically a line of cells. The \emph{alphabet} $A$ of the CA specifies the values for the states of the cells. Each cell updates its state in parallel by applying the same local rule, which is evaluated on the cell's neighborhood.

Usually, the relevant literature considers a CA as a discrete dynamical system, and studies their long-term (asymptotic) behavior which emerges from the iterated application of the local rule over multiple time steps. On the contrary, in this work we are interested in CA as mere \emph{algebraic systems}: in particular, we consider a CA as a particular kind of vectorial Boolean function. Formally, we introduce the following definition:

\begin{definition}
\label{def:ca}
Let $d,n \in \N$ such that $d \le n$, and let $b=d-1$.  A \emph{no-boundary cellular automaton} with local rule $f:\F_2^d \to \F_2$ of diameter $d$ is a vectorial Boolean function $F: \F_2^{n} \to \F_2^{n-b}$ whose $i$-th coordinate is defined as:
\begin{equation}
\label{eq:ca}
F(x_1,\cdots,x_n)_i = f(x_i,\cdots, x_{i+b})
\end{equation}
for all $i \in \{1,\cdots,n-b\}$ and $x \in \F_2^n$.
\end{definition}
Thus, for all $i \in \{1, \cdots, n-b\}$, the output coordinate $F_i$ is defined as the local rule $f$ evaluated over the neighborhood $(x_i, \cdots, x_{i+b})$. This model of CA is called ``no-boundary'' since the local rule is applied only up to the point where there are enough cells in the input to construct a $d$-variable neighborhood, that is until $i=n-b$. Therefore, the cellular array ``shrinks'' after the application of the global rule $F$, as we lose $b$ cells. However, this does not pose a problem, since as we mentiond above we are not interested in iterating the CA over multiple time steps. Therefore, we do not need to define any boundary condition (such as periodic boundaries). In what follows, we will mostly use CA to refer to the no-boundary case, specifying if other boundary conditions are used.

We now introduce orthogonal cellular automata. A \emph{Latin square} of order $N \in \N$ is a $N \times N$ square matrix $L$ whose rows and columns are permutations of $[N] = \{1,\cdots, N\}$. Thus, if we fix any row or column of $L$, all numbers from $1$ to $N$ occur exactly once. Moreover, two Latin squares $L_1,L_2$ of the same order $N$ are \emph{orthogonal} if their \emph{superposition} yields all possible pairs in the Cartesian product $[N]\times [N]$ exactly once. Orthogonal Latin squares have many applications in cryptography and coding theory, most notably for the design of secret sharing schemes and MDS codes~\cite{stinson04}.

Eloranta~\cite{eloranta93} and Mariot et al.~\cite{mariot16} independently proved that a CA defined by a bipermutive local rule can be used to define a Latin square. A local rule $f: \F_2^d \to \F_2$ is called \emph{bipermutive} if it can be written as the XOR of the leftmost and rightmost variables with a generating function of the $d-2$ central ones, i.e. $f(x_1,\cdots,x_d) = x_1 \oplus g(x_2,\cdots,x_b) \oplus x_d$, with $g: \F_2^{d-2} \to \F_2$. Then, a CA $F: \F_2^{2b} \to \F_2^b$ equipped with such a local rule $f$ corresponds to a Latin square of order $N = 2^b$. The idea is to use the left and right $b$ input cells of $F$ respectively to index the rows and the columns of a $2^b \times 2^b$ square, and then take the output of the CA as the entry of the square at those coordinates.

A pair of \emph{orthogonal CA} (OCA) is a pair of CA $F,G: \F_2^{2b} \to \F_2^b$ defined by bipermutive rules $f,g: \F_2^d \to \F_2$ such that the corresponding Latin squares of order $2^b$ are orthogonal.

\section{Related Work}
\label{sec:rel-work}
We start with an overview of the use of CA in the design of block ciphers, and especially for the construction of S-boxes. For further information, we refer the reader to the recent survey chapter on AI methods for the design of symmetric cryptographic primitives~\cite{mariot22}. 

Gutowitz~\cite{gutowitz93} was the first to propose a symmetric cryptosystem based on the iteration of both irreversible and reversible CA. The main drawback of his design was that the diffusion phase expanded the length of the ciphertext, since it was based on the preimage computation of irreversible CA.

Daemen pioneered the use of CA in cryptography through a single application of the global rule, instead of iterating it as a dynamical system. In his PhD thesis~\cite{daemen95}, he started to study a very simple local rule of diameter $d=3$, which he named $\chi$. This local rule may be succintly described as ``flip the value of the current cell if the two right neighboring cells are in the state 10''.  Using Wolfram's numbering convention~\cite{wolfram83}, which encodes the truth table vector of a local rule as a decimal number, the map $\chi$ corresponds to rule $210$. What is especially interesting about $\chi$ is that it induces an invertible CA if the length of the cellular array is odd. Moreover, Daemen et al.~\cite{daemen94} showed that the correlation and differential properties of $\chi$ are easy to analyze from an algebraic point of view, making it interesting for cryptographic applications. This rule (or a variation thereof) was adopted in several symmetric primitives designed by Daemen and other authors, such as {\sc Panama}~\cite{daemen98} and {\sc RadioGat{\'{u}}n}~\cite{bertoni06}. Most notably, the permutation in the {\sc Keccak} sponge construction~\cite{bertoni13}, which became the NIST SHA-3 standard for hash functions, uses a $5\times 5$ S-box where rule $\chi$ is applied with periodic boundary conditions. Interestingly, this S-box is the only nonlinear component in the design of {\sc Keccak}.

The research line of implementing S-boxes with a single evaluation of a CA global rule has been revived in more recent years. Picek et al.~\cite{picek17,picek17a} investigated the use of Genetic Programming (GP) to evolve S-boxes defined by CA rules. Their results showed that GP is able to discover CA-based S-boxes up to size $7\times 7$ with optimal cryptographic properties and implementation cost on par with other state-of-the-art S-boxes. Ghoshal et al.~\cite{ghoshal18} employed CA rules to define lightweight S-boxes of size $4\times 4$ that are resistant against side-channel attacks. Mariot et al.~\cite{mariot19} carried out a theoretical analysis on the cryptographic properties of S-boxes defined by CA rules, proving bounds for their nonlinearity and differential uniformity. 

Another approach which instead considers CA as a dynamical system for the synthesis of S-boxes is based on \emph{second-order CA}. The next state of a cell is computed by XORing the result of the local rule applied to its usual neighborhood with the state of the cell in the previous time step. This method ensures that the overall system is reversible. The first to investigate this approach for designing block ciphers with CA were Seredyinski et al.~\cite{seredynski04}. There, the authors used the forward evolution of a second-order CA to encrypt the whole plaintext block, rather than a small portion of it as in the SPN paradigm based on small S-boxes. Decryption then corresponded to backward evolution of the CA, granted by the second-order property. The authors performed experimental evaluations on the avalanche property of the resulting block cipher, testing over a random sample of local rules of diameter $d=5$ and $d=7$. Szaban et al.~\cite{szaban08} used the same second-order approach to define S-boxes of size $8 \times 8$, focusing on elementary local rules of diameter $d=3$. In particular, they selected the subset of rules that yielded the best values of nonlinearity and autocorrelation.

Some other worls also explored other methods to design block ciphers and S-boxes via CA, although they represent minor research threads. For instance, Marconi et al.~\cite{marconi06} studied the analogies between the Lattice Gas Automata (LGA) model for fluids and the SPN paradigm. In particular, the authors proposed to use the collision operator of the LGA model to implement the substitution layer of a block cipher.

We conclude this section with a brief outlook of the research line devoted to orthogonal cellular automata, which has been mostly investigated by the second author of this manuscript. Mariot et al. first proved in~\cite{mariot16} a necessary and sufficient condition for a pair of linear bipermutive CA to generate orthogonal Latin squares. The characterization is quite simple, since it consists in checking whether the polynomials associated to the local rules of the CA are relatively prime. This result was later developed in~\cite{mariot20} by counting the number of pairs of coprime polynomials with a nonzero constant term, and by providing a construction for maximal families of Mutually Orthogonal Latin Squares (MOLS) generated by linear bipermutive CA. These results have been subsequently used by Gadouleau et al.~\cite{gadouleau20} to devise a new construction of bent Boolean functions, which reach the highest possible nonlinearity. Later, it turned out that the construction could be greatly simplified through the formalism of linear recurring sequences, instead of using orthogonal CA~\cite{gadouleau23}. Formenti et al.~\cite{mariot22d} devised a combinatorial algorithm to enumerate all pairs of coprime polynomials with nonzero constant term, and thus all linear OCA of a given diameter. Finally, Mariot~\cite{mariot22b} considered orthogonal CA as pseudorandom generators, and devised an algorithm to compute the period of the resulting sequences when the underlying CA are linear.

The amount of theoretical results and applications developed for linear OCA contrasts with what little is known about the nonlinear setting. From a theoretical point of view, only a necessary condition on the local rules of two nonlinear OCA is currently known~\cite{mariot17}, and an inversion algorithm for the configurations of nonlinear OCA has been proposed in~\cite{mariot18}. The authors of~\cite{mariot17a} also used evolutionary algorithms to evolve pairs of nonlinear OCA. However, to date a theoretical characterization of nonlinear OCA similar to the linear case is still missing.

\section{S-boxes Generated by OCA}
\label{sec:sboxes}
We now describe our method to generate an S-box from a pair of orthogonal CA. As we said in Section~\ref{subsec:ca}, our model of CA is a particular kind of vectorial Boolean function. More precisely, given a bipermutive local rule $f: \F_2^d \to \F_2$ of diameter $d = b+1$, we can interpret the corresponding CA equipped with $f$ both as a Latin square of order $2^b$ and as a $(2b,b)$-function. However, here we are mainly interested in S-boxes for the SPN paradigm, thus we need to define an $(n,n)$-function with the same number of input and output bits. To this end, we use a pair of orthogonal CA to define the \emph{superposition S-box} as follows:
\begin{definition}
\label{def:sup-sbox}
Let $f,g: \F_2^d \to \F_2$ be two bipermutive local rules of diameter $d=b+1$ that give rise to a pair of OCA $F, G: \F_2^{2b} \to \F_2^b$. Setting $n=2b$, the \emph{superposition S-box} is the vectorial function $H: \F_2^{n} \to \F_2^{n}$ defined for all $x \in \F_2^n$ as:
\begin{equation}
\label{eq:sup-sbox}
H(x) = F(x)||G(x) \enspace ,
\end{equation}
where $||$ denotes the concatenation operator.
\end{definition}
Thus, the output of the superposition S-box is defined by concatenating the outputs of the OCA $F$ and $G$ evaluated on the same input vector. Alternatively, the first $b$ coordinates functions of $H$ correspond to the coordinates of $F$, while the last $b$ coordinates correspond to the coordinates of $G$. Hence, Equation~\eqref{eq:sup-sbox} can be explicitly rewritten as:
\begin{equation}
\label{eq:sboxoca}
H(x) = (f(x_1,\cdots,x_d),\cdots,f(x_b,\cdots,x_{n}),g(x_1,\cdots,x_d),\cdots,g(x_b,\cdots,x_n)) \enspace .
\end{equation}

A legit question is why an S-box should be defined by the superposition of two OCA, instead of using a single CA rule as done in most of the literature reviewed in Section~\ref{sec:rel-work}. As argued in~\cite{mariot22b}, where OCA are used to generate pseudorandom sequences, there are two main motivations:

\paragraph{Motivation 1: Bijectivity.} Since $F$ and $G$ are orthogonal CA, the superposition of their associated Latin squares defines a permutation over the Cartesian product $[2^b] \times [2^b]$. Remark that the set $[2^b]$ is in a one-to-one correspondence with the vector space $\F_2^b$; hence, the Cartesian product $[2^b] \times [2^b]$ is also straightforwardly mapped one-to-one onto the product space $\F_2^{b} \times \F_2^b$, which is in turn isomorphic to $\F_2^{2b}$. It follows that the superposition S-box $H$ is bijective, since it simply concatenates the output of $F$ and $G$. As we mentioned in the previous sections, the S-boxes used in SPN ciphers must be bijective, in order to ensure decryption. With a single CA, there is no such guarantee, and the cipher designer has only two alternatives: either resort to subclasses of rules for which invertibility conditions are known (as in the case of the $\chi$ map used in {\sc Keccak}~\cite{bertoni13}), or use heuristic algorithms to optimize the bijetivity of the S-box~\cite{picek17}.

\paragraph{Motivation 2: Diffusion.} The permutation defined by the superposition of two Latin squares is not a generic bijection. As a matter of fact, this permutation ``spreads'' the input in an optimal way, since taking two different ordered pairs which agree on the first (respectively, on the second) coordinate implies that their corresponding output pairs cannot have the same value on the first (respectively, on the second) coordinate. Vaudenay~\cite{vaudenay94} formalized this observation by showing that orthogonal Latin squares are $(2,2)$-multipermutations, and thus they provide an optimal diffusion between $4$-tuples formed by pairs of inputs and outputs. In the context of the superposition S-box introduced in Definition~\ref{def:sup-sbox}, this means that for all $x,x',y,y' \in \F_2^b$ such that $(x,y) \neq (x',y')$, the tuples $(x,y, F(x||y),G(x||y))$ and $(x',y', F(x'||y'),G(x'||'y))$ always disagree on at least 3 coordinates.

One might object that multipermutation-like diffusion is not a relevant criterion for the design of S-boxes, since this property is usually provided by the permutation layer in SPN ciphers rather than by the substitution layer. Moreover, diffusion layers are usually implemented with \emph{linear mappings}, such as the MDS matrix in {\sc Rijndael}~\cite{daemen20} or other similar transformations~\cite{li17}. However, there has been a recent interest also in \emph{nonlinear} diffusion layers~\cite{liu18}, which integrate confusion and diffusion. Thus, superposition S-boxes might be interesting for this specific application.

In the next section, we address the following research question concerning superposition S-boxes.

\begin{resq}
\label{rq:question}
Let $b \in \N$ and $n=2b$. Do there exist superposition S-boxes $H: \F_2^{n} \to \F_2^{n}$ defined as in Equation~\eqref{eq:sup-sbox} that are nonlinear? And if they exist, do they achieve a high nonlinearity?
\end{resq}

In this work we focus only on nonlinearity, as this is one of the most important cryptographic properties when considering S-boxes for the design of block ciphers. By ``high nonlinearity'', we mean a nonlinearity value close to the theoretical upper bounds, for which we refer the reader to~\cite{carlet21}.

\section{Exhaustive Search Experiments}
\label{sec:search}
As we saw in Section~\ref{sec:rel-work}, there is a rich theory pertaining linear OCA. Thus, at a first glance one might be tempted to exploit these results to construct superposition S-boxes. Unfortunately, the S-box $H$ associated to two linear OCA is also linear: indeed, any linear combination of linear coordinates will always yield a linear component function. Therefore, superposition S-boxes generated by linear OCA are certainly useless in the design of substitution layers in SPN ciphers.

For this reason, we considered S-boxes generated by nonlinear OCA. Since there is no theoretical characterization of nonlinear OCA yet, the most natural way to address this problem is to exhaustively generate all pairs of bipermutive rules of diameter $d$, check if the resulting Latin squares are orthogonal, and in that case compute the nonlinearity of the associated S-box.

The space of all Boolean functions of $d$ variables is $2^{2^d}$, since one needs to define a $2^d$-bit truth table to uniquely identify a function. However, we are only interested in bipermutive local rules, which are $2^{2^{d-2}}$ for a given diameter $d$. This is due to the fact that the truth table of a bipermutive function is determined only by the central $d-2$ variables, since the leftmost and rightmost ones are always XORed. Hence, to exhaustively enumerate all pairs of bipermutive local rules of diameter $d$, one needs to generate $2^{2^{d-2}} \cdot 2^{2^{d-2}} = 2^{2^{d-1}}$ elements in total.

We adopted the naive enumeration approach outlined above in the conference version of this paper~\cite{mariot22}, applying it to the enumeration all nonlinear OCA pairs of diameter $d=4$ and $d=5$ (therefore, superposition S-boxes of sizes $6\times 6$ and $8 \times 8$, respectively)\footnote{We discarded the case $d=3$ since there are no nonlinear OCA pairs for that diameter.}. At that point, the main finding of our search was that \emph{all superposition S-boxes of those sizes turned out to be linear}. This was quite surprising, since one could reasonably expect that at least some of the nonlinear OCA pairs would give rise to a nonlinear superposition S-box. Instead, the main conclusion of our experiments was that even in the case where the OCA are defined by nonlinear local rules, there is always at least one nontrivial combination of (nonlinear) coordinate functions that results in an affine function. 

Clearly, this finding should be corroborated with further experiments on nonlinear OCA defined by larger local rules. In principle, one could still employ the naive enumeration algorithm for diameter $d=6$, since in this case one has to generate $2^{2^{6-1}} \approx 4.3 \cdot 10^9$ pairs, which is still computationally feasible. However, we employed the more efficient combinatorial algorithm described in~\cite{mariot17}, which enumerates only pairs of nonlinear bipermutive rules that are \emph{pairwise balanced}. As proved in the same paper, pairwise balancedness is a necessary condition for two bipermutive rules to generate a pair of OCA. Thus, the number of pairs to visit reduces to about $8.4 \cdot 10^8$. Using this algorithm, one obtains $532800$ nonlinear OCA pairs of diameter $d=6$\footnote{Notice that this number can be further divided by 8, thus giving $66600$ pairs. This is due to the fact that there are three symmetry classes that preserve the orthogonality property of an OCA pair, as shown in~\cite{mariot17}. However, in this work we stick to the total number of pairs.}.

We thus focused on this set of rules and generated the corresponding superposition S-boxes, computing their nonlinearity. Contrary to the evidence gathered for  $d=4$ and $d=5$, \emph{we found nonlinear S-boxes for diameter $d=6$}. Therefore, this finding allows us to answer affirmatively to the first part of Research Question~\ref{rq:question}, and falsifies one of the natural hypotheses that arised after seeing what happens for lower diameters: namely, that the superposition S-boxes generated by OCA are always linear, independently of the nonlinearity of the underlying local rules.

Table~\ref{tab:lcs} summarizes our exhaustive search experiments for each considered diameter $d$. In particular, we further classified linear S-boxes with respect to the dimension of their Linear Components Space (LCS). Column $nl(H)$ denotes the nonlinearity of the superposition S-boxes, $\#nl(H)$ the number of S-boxes attaining that nonlinearity, $dim$ their LCS dimension, and $\#dim$ the number of S-boxes whose LCS have that dimension.
\setlength{\tabcolsep}{0.8em}
\begin{table}[t]
	\centering
	\caption{Classification of OCA-based S-boxes of diameter $4 \le d \le 6$ in terms of their nonlinearity and LCS dimensions.}
	\begin{tabular}{ccccc}
		\hline\noalign{\smallskip}
		$d$ & $nl(H)$ & $\#nl(H)$ & $dim$ & $\#dim$ \\
		\noalign{\smallskip}\hline\noalign{\smallskip}
		$4$ & $0$ & $32$ & $3$ & $32$ \\
		\hline\noalign{\smallskip}
        \multirow{2}{*}{$5$} & \multirow{2}{*}{$0$} & \multirow{2}{*}{$1536$} & $3$ & $64$ \\
						   \cline{4-5}\noalign{}
                           &                        &                         & $4$ & $1472$ \\
		\hline\noalign{\smallskip}
        \multirow{8}{*}{$6$} & $128$ & $4448$ & $0$ & $4448$ \\
                             & $64$  & $64$ & $0$ & $64$ \\
                             \cline{2-5}\noalign{}
                             & \multirow{5}{*}{$0$} & \multirow{5}{*}{$528288$} & $3$ & $384$ \\
                             \cline{4-5}\noalign{}
                             &                      &                           & $4$ & $1984$ \\
                             \cline{4-5}\noalign{}
                             &                      &                           & $5$ & $525920$ \\
        \hline\noalign{\smallskip}                             
	\end{tabular}
	\label{tab:lcs}
\end{table}

One can see from the table that the great majority of $d=6$ OCA pairs still generate S-boxes of size $10 \times 10$ that are linear. Moreover, the remaining S-boxes have a very low value of nonlinearity, i.e. either $64$ or $128$. To give a reference, the \emph{Sidelnikov-Chabaud-Vaudenay} bound~\cite{carlet21} states that the upper bound for the nonlinearity of a $(10,10)$-function is 480. Moreover, S-boxes of size $10 \times 10$ are seldom used in real-world ciphers. Therefore, we can confirm the conclusion of our conference paper~\cite{mariot22}: \emph{nonlinear OCA cannot be used to design S-boxes in the substitution layer of SPN ciphers}. Still, we remark that these S-boxes might be useful in the design of nonlinear diffusion layers~\cite{liu18}, since it is not required to reach a nonlinearity close to the theoretical upper bounds in that use case. Indeed, the goal of a nonlinear diffusion layer is to provide extra confusion in addition to that already given by a classic substitution layer.

\section{The Linear Component Space of OCA S-boxes}
\label{sec:lcs}
The computer search presented in the previous section shows that most of the superposition S-boxes obtained by nonlinear OCA are linear. Therefore, their linear components spaces are nontrivial, or equivalently they have a strictly positive dimension. We now analyze the LCS of these S-boxes more in detail, uncovering an interesting coding-theoretic structure. Below, we recall some basic facts about linear error correcting codes. Some good references that the reader may consult for a more complete overview of the topic are~\cite{macwilliams77} and~\cite{mceliece02}.

A $(n,k)$-\emph{binary linear code} $C \subseteq \F_2^n$ of length $n \in \N$ and dimension $k \le n$ is a $k$-dimensional subspace of $\F_2^n$. The vectors in $C$ are also called the \emph{codewords}, and they are characterized by a \emph{minimum Hamming distance} $d_{min}$, which determines how many errors the code can correct. Suppose that $\{b_1, b_2, \cdots, b_k\} \subseteq C$ is a set of $k$ linearly independent vectors in $C$, i.e. they form a basis of the subspace. Then, these vectors can be seen as the rows of a $k\times n$ \emph{generator matrix} $G$ of the code $C$. The encoding of a $k$-bit message $m \in \F_2^k$ into an $n$-bit codeword $c \in C$ is given by the multiplication of $m$ by the generator matrix, i.e. $c = mG$. A \emph{parity-check matrix} $P$ for $C$ is a $n \times k$ matrix such that $y\cdot P = \underbar{0}$ if and only if $y \in C$. The parity-check matrix is used in the decoding step, and the $k$-bit vector resulting from the multiplication of a $n$-bit message by $P$ is also called the \emph{syndrome}, which is employed to correct errors occured during transmission.

Here, we focus on a particular type of linear codes, namely \emph{polynomial codes}. These are defined by a particular basis that is obtained by shifting the coefficients of a \emph{generator polynomial}. Let $g(X) = a_1 + a_2X + \cdots + X^{t} \in \F_2[X]$ be a polynomial of degree $t \le n$. Then, the generator matrix $G$ of the associated $(n,k)$ polynomial code $C$ is defined as:
\begin{equation}
\label{eq:ca-matr-f}
G = 
\begin{pmatrix}
a_0    & \cdots & a_{t-1} & 1 & 0 & \cdots & \cdots & \cdots & \cdots & 0 \\
0      & a_0    & \cdots  & a_{t-1} & 1 & 0 & \cdots & \cdots & \cdots & 0 \\
\vdots & \vdots & \vdots & \ddots  & \vdots & \vdots & \vdots & \ddots & \vdots & \vdots \\
0 & \cdots & \cdots & \cdots & \cdots & 0 & a_0 & \cdots & a_{t-1} & 1 \\
\end{pmatrix} \enspace .
\end{equation}
Thus, for all $i \in \{1,\cdots,k\}$ the $i$-th row of the matrix is obtained by shifting the coefficients of $g$ by $i-1$ positions to the right. Further, a polynomial code is \emph{cyclic} if and only if its generator polynomial $g$ divides $X^{n}+1$\footnote{Notice that certain authors (see e.g. Kasami et al.~\cite{kasami68}) use the term \emph{polynomial code} to actually refer to a \emph{subclass} of cyclic codes. Here, instead, we follow Gilbert and Nicholson's notation (see~\cite{gilbert04}), where a polynomial code is a generalization of a cyclic code (specifically, the generator polynomial does not need to be a divisor of $X^n+1$)}. In this case, the resulting code is closed under \emph{cyclic shifts}, that is, if $c = (c_1, \cdots, c_{n-1}, c_n) \in C$ then $c' = \sigma(c) = (c_n, c_1,\cdots, c_{n-1})$ also belongs to $C$.

On the other hand, in a generic polynomial code not all codewords are necessarily obtained as (non-cyclic) shifts. But clearly, at least each member of the basis that forms the generator matrix in~\eqref{eq:ca-matr-f} is either a right shift (which we denote by $\sigma_R$) or a left shift (denoted by $\sigma_L$) of another member of the basis.

In the conference version of our paper~\cite{mariot22} we analyzed the LCS spaces of the superposition S-boxes for diameters $d=4$ and $d=5$ and found out that \emph{all of them are polynomial codes}. We conjectured that this is true for any diameter, i.e. if the superposition S-box $H: \F_2^{n} \to \F_2^n$ generated by two OCA is linear, then its LCS is always a polynomial code, for any $d \in \N$. As a first check, here we extended our empirical investigation to the case of diameter $d=6$, and observed that the conjecture still held.

We now prove the above conjecture. First, we need the following auxiliary lemma about the linear components of a single no-boundary CA. We state this result without proof, since it is a trivial generalization of Lemma 1 in~\cite{mariot22e}\footnote{In particular, Lemma 1 in~\cite{mariot22e} considers only the specific case of the component function $v\cdot F$ where $v = (1,\cdots,1)$, i.e. the linear combination that sums all coordinates. But one can straightforwardly prove the same result for any other component function by following the same structure of the proof.}.
\begin{lemma}
\label{lm:eq-deg}
Let $f: \F_2^d \to \F_2$ be a Boolean function of $d$ variables, $n \ge d$ and $F: \F_2^n \to \F_2^{n-d+1}$ the no-boundary CA of length $n$ equipped with $f$ as a local rule. Then, for any $v \in \F_2^{n-d+1} \setminus \{\underbar{0}\}$, the algebraic degree of the component function $v \cdot F$ equals the algebraic degree of $f$.
\end{lemma}
A consequence of Lemma~\ref{lm:eq-deg} is that if the local rule $f$ is nonlinear, then any component function of the CA $F$ must also be nonlinear, since it must have degree greater than 1. We now use this remark to prove the following result:
\begin{lemma}
\label{lm:part}
Let $F,G: \F_2^{2b}\to \F_2^b$ be two OCA respectively defined by two nonlinear bipermutive rules $f,g: \F_2^d \to \F_2$ of diameter $d=b+1$, and consider the superposition S-box $H: \F_2^{n} \to \F_2^{n}$ defined by $F$ and $G$ with $n=2b$. If $v \in \F_2^{n} \setminus \{\underbar{0}\}$ is such that $nl(v\cdot H) = 0$, then the support of $v$ cannot contain only indices less than or equal to $b$, or only indices greater than or equal to $b+1$.
\begin{proof}
By contradiction, assume that $supp(v) = \{i_1,i_2,\cdots, i_l\}$ and $i_j \le b$ for all indices $j \in \{1,\cdots, l\}$. This means that the XOR of the coordinate functions $H_{i_1}, H_{i+2},\cdots, H_{i_l}$ gives an affine function. By definition of $H$, this sum corresponds to the following component function of $F$:
\begin{equation}
v'\cdot F(x) = f(x_{i_1},\cdots,x_{i_1+b}) \oplus f(x_{i_2},\cdots,x_{i_2+b}) \oplus \cdots \oplus f(x_{i_l},\cdots,x_{i_l+b}) \enspace .
\end{equation}
Hence, we have that $nl(v'\cdot F) = 0$. But this contradicts Lemma~\ref{lm:eq-deg}, since we assumed that $f$ is nonlinear. Therefore, any component function of $F$ (or equivalently, any component of $H$ whose right half is zero) must be nonlinear as well.
A symmetric reasoning stands by assuming that $i_j \ge b+1$ for all $j \in \{1,\cdots,l\}$, using the nonlinearity of $g$.
\end{proof}
\end{lemma}
Lemma~\ref{lm:part} tells us that if a component function $v\cdot H$ of the superposition S-box $H$ is affine, then the support of $v$ must be spread in both halves of the vector. The next lemma shows that if we have a linear component $v\cdot H$ where the two halves of $v$ are such that at least their last or first coordinate is zero, then the right (respectively, left) shift of $v$ also gives a linear component of $H$:
\begin{lemma}
\label{lm:shift}
Given $n=2b$, suppose that $H: \F_2^{n} \to \F_2^n$ is a linear S-box defined by two OCA $F,G: \F_2^{2b} \to \F_2^b$ equipped by nonlinear bipermutive rules $f,g: \F_2^{d} \to \F_2$ of diameter $d=b+1$. Further, assume that $v = (v_1,\cdots, v_b, v_{d}, \cdots v_n) \in \F_2^{n} \setminus \{\underbar{0}\}$ is such that $nl(v \cdot H) = 0$, and that at least $v_b$ and $v_n$ (respectively, $v_1$ and $v_d$) are zero. Then, the vector defined as the right (respectively, left) shift of $v$ is such that $nl(v'\cdot H) = 0$.
\begin{proof}
By Lemma~\ref{lm:part} we know that $supp(v)$ is such that the subsets $\{i: i\le b, v_i \neq 0\}$ and $\{i: i\ge d, v_i \neq 0\}$ are both nonempty. Therefore, we can write the component function $v\cdot H$ as follows:
\begin{equation}
\label{eq:sum}
    v \cdot H = \bigoplus_{i \in supp(v), \ i \le b} f(x_{i},\cdots,x_{i+b}) \bigoplus_{i \in supp(v), \ i \ge d} g(x_{i-b}, \cdots, x_i) \enspace .
\end{equation}
Since $nl(v\cdot H) = 0$, it means that all nonlinear terms in Equation~\eqref{eq:sum} (i.e., those of degree greater than 1) cancel out, and only the affine terms remain. Suppose now that $v_b=v_n=0$. Then, the component $v'\cdot H$ obtained by shifting right the coordinates of $v$ is equal to:
\begin{equation}
\label{eq:sum-1}
    v \cdot H = \bigoplus_{i \in supp(v), \ i \le b} f(x_{i+1},\cdots,x_{i+b+1}) \bigoplus_{i \in supp(v), \ i \ge d} g(x_{i-b+1}, \cdots, x_{i+1}) \enspace .
\end{equation}
As it can be seen in Equation~\eqref{eq:sum-1}, the linear combination has the same structure of~\eqref{eq:sum}, with the only difference that all indices are increased by 1. But then, it follows that the nonlinear terms of~\eqref{eq:sum-1} must also cancel out. Therefore, one has that $nl(v'\cdot H) = 0$. The case where $v_1=v_d=0$ is symmetrical, by considering the left shift of $v$.
\end{proof}
\end{lemma}
Thus, we proved that if a component function belongs to the LCS of a superposition S-box $H$ and there is ``enough space'' either on the left or on the right, then its left (respectively, right) shift also belongs to the LCS of $H$. Remark that such components always exist: indeed, even if one takes two linear components that cannot be shifted (because they both have 1 on the leftmost and the rightmost coordinates, for instance), then their sum can. Therefore, the LCS of a superposition S-box cannot be made entirely of vectors that cannot be shifted. Further, observe that if we construct a family of linear components by consecutive shifts, then they are all linearly independent, and the largest family that can be obtained in this way constitutes a basis for the LCS. We have thus finally proved our main result:
\begin{theorem}
\label{thm:pol-cod}
Given $n=2b$, let $H: \F_2^n \to \F_2^n$ be a linear S-box defined by two nonlinear OCA $F,G: \F_2^{2b} \to \F_2^b$. Then, the LCS of $H$ is a polynomial code.
\begin{proof}
By the remark above, one can construct a basis of the LCS of $H$ by consecutively shifting one linear component and iteratively applying Lemma~\ref{lm:shift}. The vectors of this basis can be used to form a generator matrix of the form of Equation~\eqref{eq:ca-matr-f}.
\end{proof}
\end{theorem}

We conclude this section by providing a classification of thelinear S-boxes obtained for diameter $4 \le d \le 6$ in terms of their generator polynomials, which uniquely identify the respective LCS.

Referring to Table~\ref{tab:lcs}, all the $32$ LCS for diameter $d=4$ and the $768$ LCS for $d=5$ and nonlinearity $(4,4)$ are actually $(2b,b)$ cyclic codes with generator $g(X) = 1 + X^b$. For diameter $d=5$ and nonlinearity $(8,8)$, the $704$ S-boxes with LCS of dimension $4$ are again $(2b,b)$ cyclic codes with generator $1 + X^b$, while the remaining $64$ are split in four classes, each of size $16$, defined by the following generators:
\begin{align}
\nonumber
X + X^4 + X^5; \enspace\enspace &1 + X^4 + X^5;  \\
\nonumber
1 + X + X^4; \enspace\enspace &1 + X + X^6.  \\
\end{align}

For diameter $d=6$, we have a total of 528288 linear S-boxes. In this case, the great majority of LCS have dimension $5$, and they are again all defined by the generator polynomial $g(X) = 1 + X^b$. For dimension $4$, there are $4$ classes each composed of $496$ elements, represented by the following polynomials:
\begin{align}
\nonumber
X + X^5 + X^6; \enspace\enspace &1 + X + X^6;  \\
\nonumber
1 + X^5 + X^6; \enspace\enspace &1 + X + X^5.  \\
\end{align}

Finally, for dimension $3$ we found 14 classes, two of which contain $96$ elements and are represented respectively by the polynomials:
\begin{align}
\nonumber
1 + X + X^2 + X^5 + X^7; \enspace\enspace &1 + X^2 + X^5 + X^6 + X^7; 
\end{align}
The remaining 12 classes, on the other hand, all have 16 elements and are represented by the following polynomials:
\begin{align}
\nonumber
X^2 + X^5 + X^7; \enspace\enspace &X^2 + X^5 + X^6 + X^7;  \\
\nonumber
X + X^2 + X^5 + X^6 + X^7; \enspace\enspace &1 + X + X^2 + X^5 + X^6;  \\
\nonumber
1 + X^5 + X^7; \enspace\enspace &1 + X^2 + X^5;  \\
\nonumber
1 + X + X^2 + X^6 + X^7; \enspace\enspace &1 + X + X^5 + X^6 + X^7;  \\
\nonumber
1 + X^5 + X^6 + X^7; \enspace\enspace &1 + X + X^2 + X^5;  \\
\nonumber
1 + X + X^2 + X^7; \enspace\enspace &1 + X^2 + X^7;  \\
\end{align}

A final interesting remark is that the generator polynomial $1 + X^b$ (which accounts for the great majority of the LCS examined here) corresponds to the case where the local rules $f$ and $g$ \emph{share the same nonlinear terms in their ANF}. Indeed, this is the only way that the linear components of the form $F_i \oplus G_i$ for $i \in \{1,\cdots b\}$ can collapse to an affine function, since $f$ and $g$ are evaluated on the same neighborhood.

\section{Conclusions}
\label{sec:outro}
In this paper, we investigated S-boxes defined by pairs of orthogonal cellular automata. The motivation for considering this particular approach resides in the bijectivity of the resulting S-box, which is ensured by the orthogonality of the Latin squares generated by the OCA, and the diffusion property granted by the fact that orthogonal Latin squares are $(2,2)$-multipermutations. We extended our search experiments up to diameter $d=6$ and confirmed the practical implications that we introduced in the conference version of this paper~\cite{mariot22}: OCA S-boxes cannot be used to design the substitution layer of a SPN block cipher, since most of them are linear, and the few nonlinear ones have too small nonlinearity values. Still, we believe that OCA S-boxes might be of interest for the design of nonlinear diffusion layers~\cite{liu18}, where the goal is not to reach the highest possible nonlinearity, but rather to add some extra confusion in the permutation layer. Future research could focus on this aspect to investigate further the practical applicability of nonlinear OCA.

Moreover, we settled the conjecture formulated in~\cite{mariot22} by formally proving that if an S-box generated by a pair of nonlinear OCA is linear, then its linear component space is a polynomial code. We finally extended the classification of these LCS in terms of their generator polynomials up to diameter $d=6$.

An interesting consequence of Theorem~\ref{thm:pol-cod} stems from the fact that polynomial codes are actually the images of linear cellular automata. Indeed, the generator matrix in Equation~\eqref{eq:ca-matr-f} corresponds to the transition matrix of a linear CA. Therefore, what we proved is that the LCS of a linear S-box generated by a pair of OCA is itself a cellular automaton. We deem this discovery potentially interesting for future research on the theoretical characterization of nonlinear OCA pairs, which is still an open problem.

\subsection*{Appendix: Source Code and Experimental Data}
The source code and experimental data are available at \url{https://github.com/rymoah/orthogonal-ca-sboxes}.

\bibliographystyle{abbrv}
\bibliography{bibliography}

\end{document}